\documentclass[12pt]{article} 
\usepackage{amssymb,amsmath}
\usepackage[noblocks]{authblk}
\usepackage[top=0.75in, bottom=0.75in, left=0.75in, right=0.75in, dvips]{geometry}
\usepackage{caption}
\date{}

\begin{document}
\textwidth 10.0in 
\textheight 9.0in 
\topmargin -0.60in
\title{Classical Mechanics}
\author[1,2]{D.G.C. McKeon\thanks{Email: dgmckeo2@uwo.ca}}
\affil[1] {Department of Applied Mathematics, The
University of Western Ontario, London, ON N6A 5B7, Canada}
\affil[2] {Department of Mathematics and
Computer Science, Algoma University, Sault St.Marie, ON P6A
2G4, Canada}
\maketitle

\maketitle

\section*{ Classical Mechanics}

These notes provide an introduction to a number of those topics in Classical Mechanics that are useful for field theory.

\subsection*{\large 1.1 Lagrange's Equations and The Action Principle}

     In analysing a physical system, the principal goal is to study its time evolution. If the state of a system is known at time t=0, we wish to know how it will appear at subsequent times. For Newton, this meant that the positions $\textbf{r}_i(t)$ and velocities $\textbf{v}_i(t)=\dot{\textbf{r}}_i(t)$ of a collection of particles $i=1,...,N$ would evolve in time according to his famous equation $\textbf{F}=m\textbf{a}$. Since $\textbf{a}(t)$, the acceleration, is the time derivative of the velocity, this is a second order equation and hence one needs to have both the initial positions and velocities of all particles to see what happens to a system as times elapses.
     
     For the particular case of a conservative force field, the force $\textbf{F}$ can be written as the gradient of a time independent and velocity independent scalar function $V$, the potential, so that Newton's equation reduces to
\begin{equation}
m\ddot{\textbf{r}}=-\nabla V.
\end{equation}
Upon multiplying this equation by $\dot{\textbf{r}}$ , this equation can be integrated once with respect to time to yield the result
\begin{equation}
E=\frac{1}{2}m\dot{\textbf{r}}^2+V\equiv T+V
\end{equation}
where $E$, the energy, is a conserved quantity. (Remarkably, Newton himself didn't use conserved quantities in his analysis of dynamical systems; he worked entirely from the equations of motion.)
     
     In passing to the Lagrangian formalism, one makes use of generalized coordinates $q_i(t)$. The space of these coordinates is generally called ``configuration space''. It is easily shown that if $\textbf{r}_i=\textbf{r}_i(q_1...q_n)$, then Newton's equation can be rewritten in the form  
\begin{equation}
\frac{d}{dt}\frac{\partial L}{\partial \dot{q}_i}-\frac{\partial L}{\partial q_i}=0
\end{equation}
where $L=L(q,\dot{q})$, the Lagrangian, is $T-V$. (Including explicit time dependence in $L$ is trivial.)

In order to show that eqs. (1) and (3) are equivalent, one first notes that eq. (1) can be written as 
\begin{equation}
\frac{d}{dt} \frac{\partial T}{\partial \dot{r}_i} + \frac{\partial V}{\partial r_i} =0.
\end{equation}
Then, since $\dot{r}_i = \frac{\partial r_i}{\partial q_j} \dot{q}_j$ implies that $\frac{\partial \dot{r}_i}{\partial \dot{q}_j} = \frac{\partial r_i}{\partial q_j}$ 
we see that
\begin{equation}
\frac{\partial T}{\partial \dot{q}_i} = \frac{\partial T}{\partial \dot{r}_j} \frac{\partial \dot{r}_j}{\partial \dot{q}_i} = 
\frac{\partial T}{\partial \dot{r}_j} \frac{\partial r_j}{\partial q_i}.
\end{equation}
Consequently, we have
\begin{equation}
\frac{d}{dt} \frac{\partial T}{\partial \dot{q}_i} = \left(\frac{d}{dt} \frac{\partial T}{\partial \dot{r}_j} \right) 
\frac{\partial r_j}{\partial q_i} + \frac{\partial T}{\partial \dot{r}_j} \frac{\partial \dot{r}_j}{\partial q_i}
\end{equation}
which by eq. (4) becomes
\begin{equation}
= - \frac{\partial V}{\partial r_j} \frac{\partial r_j}{\partial q_i} + \frac{\partial T}{\partial q_i}\\
= \frac{\partial}{\partial q_i} (T - V).
\end{equation}
This is equivalent to eq. (3).

The solution of eq. (3) can be seen to extremize the action integral
\begin{equation}
S=\int^{t_2}_{t_1}L(q_i(t),\dot{q}_i(t))dt.
\end{equation}
To show this, let
\begin{equation}
q_i(t)=q_{\textsc{cl}i}(t)+\epsilon  \delta q_i(t).
\end{equation}
with $\delta q_i(t_1)=\delta q_i(t_2)=0$. (The subscript \textsc{cl} refers to the ``classical'' trajectory.) We then have the action $S$ depending on the parameter $\epsilon$. The requirement that $\frac{dS}{d\epsilon}$ vanishes when $\epsilon$ equals zero leads to  
\begin{equation}
\int^{t_2}_{t_1}\left(\frac{\partial L(q_\textsc{cl}(t),\dot{q}_\textsc{cl}(t))}{\partial q_i(t)}-\frac{d}{dt}\frac{\partial L(q_\textsc{cl}(t),\dot{q}_\textsc{cl}(t)}{\partial \dot{q}_i(t)}\right)\delta q_i(t)dt=0.        
\end{equation}       
If this were to vanish for arbitrary $\delta q(t)$, then $q_\textsc{cl}(t)$ would have to satisfy the Lagrange eq. (3). 
     
    When dealing with this second order differential equation, one normally specifies the initial values of $q_i$ and $\dot{q}_i$; in the variational approach, the initial and final values of $q_i$ are fixed. In both cases, two boundary conditions are applied for each value of $i$.

     In principle, the Lagrangian could depend on derivatives of $q_i(t)$ beyond the first. If, say, $L=L(q_i(t),\dot{q}_i(t),\ddot{q}_i(t))$, then the integral $\int^{t_2}_{t_1}Ldt$ would have an extremum on the trajectory satisfying the fourth order equation 
\begin{equation}
\frac{\partial L}{\partial q_i}-\frac{d}{dt}\frac{\partial L}{\partial \dot{q}_i}+\frac{d^2}{dt^2}\frac{\partial L}{\partial \ddot{q}_i}=0
\end{equation}
provided that the initial and final values of $q_i$ and $\dot{q}_i$ are specified. This extension of Lagrange's equations was first considered by Ostrogradsky.

\subsection*{\large 1.2 Hamilton's Equations and Poisson Brackets}

    In the Lagrangian approach to dynamics, the only independent variables are the $N$ components of $q_i$. If $L$ depends only on $q_i$ and $\dot{q}_i$, then the time evolution of $q_i$ is dictated by $N$ second order coupled ordinary differential equations. It is possible in this case to introduce a further $N$ independent variables $p_i$ (the canonical momenta) and through a Legendre transformation arrive at a set of $2N$ first order ordinary differential equations in the 2N dimensional ``phase space'' of the $q$'s and $p$'s that determines their time evolution.
    
    One begins by defining 
\begin{equation}
p_i=\frac{\partial L}{\partial \dot{q}_i}
\end{equation} 
and then introduces a Hamiltonian $H$ using the Legendre transformation
 \begin{equation}
 H(q_i,p_i)=\sum_{i} p_i\dot{q}_i-L(q_i,\dot{q}_i).
 \end{equation}
(We henceforth will use the Einstein summation convention for repeated indices.)
It is understood for the present that we are able to eliminate dependence on the $\dot{q}_i$ by expressing them in terms of the $p_i$. It follows that, $H$ does not have dependence on $\dot{q}_i$, as $\frac{\partial H}{\partial \dot{q}_i}$ vanishes on account of the way in which $p_i$ and $H$ have been defined in eqs. (12) and (13).
    Variation of $H$ can be expressed in two ways; assuming that $H$ has no explicit time dependence, first we see that 
\begin{equation}
dH=\frac{\partial H}{\partial q_i}dq_i+\frac{\partial H}{\partial p_i}dp_i
\end{equation}
and then from the definition of $H$
\begin{equation}
dH=\dot{q}_idp_i+p_id\dot{q}_i-\frac{\partial L}{\partial q_i}dq_i-\frac{\partial L}{\partial \dot{q}_i}d\dot{q}_i.
\end{equation}
In eq. (11) we can replace $\frac{\partial L}{\partial \dot{q}}$ by $p$ (from the definition of $p$) and $\frac{\partial L}{\partial{q}}$ by $\dot{p}$ (from Lagrange's equations). Matching coefficients of $dp$ and $dq$ in eqs. (14) and (15) then yield the $2N$ Hamilton's equations
\begin{equation}
\dot{q}_i=\frac{\partial H}{\partial p_i}
\end{equation}
\begin{equation}
\dot{p}_i=-\frac{\partial H}{\partial q_i}
\end{equation}
   
    We first note that if $L$, and hence $H$, has explicit time dependence, then
\begin{equation}
\frac{dH}{dt}=\frac{\partial H}{\partial q_i}\dot{q}_i+\frac{\partial H}{\partial p_i}\dot{p}_i+\frac{\partial H}{\partial t}.
\end{equation}
On account of the Hamilton's equations of motion, $H$ is a constant in time provided it has no explicit time dependence. If 
\begin{equation}
L=\frac{1}{2}M_{i,j}\dot{q}_i\dot{q}_j-V(q_i),
\end{equation}
then it follows that $H=T+V$, numerically equal to the total energy $E$ when the equations of motion are satisfied, with $T = \frac{1}{2} M_{i,j} \dot{q}_i\dot{q}_j$ 
being the kinetic energy.      
    
    Hamilton's equation can in fact be derived by requiring that the action integral
\begin{equation}
S=\int^{t_2}_{t_1}(p_i\dot{q}_i-H(q_i,p_i))dt
\end{equation}
be an extremum under independent variations of both $q_i$ and $p_i$, provided that $q_i$ is fixed at the initial and final times; $p_i$ need not be fixed in this way.

    It is convenient to define the Poisson bracket of two dynamical variables $A(q,p)$ and $B(q,p)$ to be
 \begin{equation}
 \left\{A,B\right\}_{PB}=\left(\frac{\partial A}{\partial q_i}\frac{\partial B}{\partial p_i}-\frac{\partial B}{\partial q_i}\frac{\partial A}{\partial p_i}\right).
 \end{equation}
(The subscript ``$PB$'' will henceforth be written explicitly only if there is an ambiguity.)

It follows from Hamilton's equations that 
\begin{equation}
\frac{dA(q(t),p(t))}{dt}=\frac{\partial A}{\partial q_i}\dot{q}_i+\frac{\partial A}{\partial p_i}\dot{p}_i
\end{equation}
can be written
\begin{equation}
\frac{dA}{dt}=\left\{A,H\right\}.
\end{equation}
This equation allows us to reduce much of dynamics to an algebraic problem as the Poisson brackets satisfy some general relations that follow from their definitions. We note that:
\begin{equation}
i- \left\{q_i,q_j\right\}=0=\left\{p_i,p_j\right\} 
\end{equation}
\begin{equation}
ii- \left\{q_i,p_j\right\}=\delta_{i,j}
\end{equation}
\begin{equation}
iii-\left\{A,B\right\}=-\left\{B,A\right\}
\end{equation}
\begin{equation}
iv-\left\{A+B,C\right\}=\left\{A,C\right\}+\left\{B,C\right\}
\end{equation}
\begin{equation}
v-\left\{A,BC\right\}=B\left\{A,C\right\}+\left\{A,B\right\}C
\end{equation}
\begin{equation}
vi-\left\{A,\left\{B,C\right\}\right\}+\left\{B,\left\{C,A\right\}\right\}+\left\{C,\left\{A,B\right\}\right\}=0.
\end{equation}
Only property $\textit{vi}$, the Jacobi identity, is non-trivial to prove.
From this identity, it follows that if $A$ and $B$ both have vanishing Poisson brackets with the Hamiltonian, then so does $\left\{A,B\right\}$.

    The canonical formalism can also be developed to deal with the situation in which the Lagrangian depends on $\ddot{q}$ as well as $q$ and $\dot{q}$. In this case, we make the definitions $v=\dot{q}$, $p=\frac{\partial L}{\partial \dot{q}}-\frac{d}{dt}\left(\frac{\partial L}{\partial \ddot{q}}\right)$ and $\pi =\frac{\partial L}{\partial
\ddot{q}}$; eq. (11) can then be expressed as a set of first order equations $\dot{q}=\frac{\partial H}{\partial p}$, $\dot{v}=\frac{\partial H}{\partial \pi}$, $\dot{p}=-\frac{\partial H}{\partial q}$ and $\dot{\pi}=-\frac{\partial H}{\partial v}$ where the Hamiltonian is a function of the independent variables 
$q$, $v$, $p$ and $\pi$ 
given by $H(q,v;p,\pi )=p\dot{q}+\pi \dot{v}-L(q,\dot{q},\ddot{q})$.

\subsection*{\large 1.3 Canonical Transformations and the Hamilton-Jacobi Equation}

    A change of variables from $(q_i,p_i)$ to $(Q_i,P_i)$ is said to be ``canonical'' if it leaves the form of Hamilton's equations unchanged. Thus if $(q_i,p_i)$ satisfy Hamilton's eqs. (eqs. (16) and (17)), there exists a function $K(Q_i,P_i)$ such that 
\begin{equation}
\dot{Q}_i=\frac{\partial K}{\partial P_i}
\end{equation}
\begin{equation}
\dot{P}_i=-\frac{\partial K}{\partial Q_i}.
\end{equation}   

    The action $S=\int^{t_2}_{t_1}(P_i\dot Q_i-K)$ is extremized by the solution to these equations, just as the action of eq. (20) leads to Hamilton's equations for $q_i$ and $p_i$. As the variations of $(q,p)$ and $(Q,P)$ vanish at the end points $t_1$ and $t_2$, these two integrands can differ at most by the time derivative of some function $F$, so that
\begin{equation}
p_i\dot{q}_i-H=P_i\dot{Q}_i-K+\frac{dF}{dt}.
\end{equation}
     
The relation $F$ can be function of only $2N$ of the $4N$ variables $(q_i,p_i,Q_i,P_i)$; we initially examine
\begin{equation}
F=F_1(q_i,Q_i,t).
\end{equation} 
(We now consider the possibility that $H$ has explicit time dependence.) As 
\begin{equation}
\frac{dF}{dt}=\frac{\partial F_1}{\partial q_i}\dot{q}_i+\frac{\partial F_1}{\partial Q_i}\dot{Q}_i+\frac{\partial F_1}{\partial t}
\end{equation}
together eqs. (32) and (34) imply that 
\begin{equation}
p_i=\frac{\partial F_1}{\partial q_i}
\end{equation}
\begin{equation}
P_i=-\frac{\partial F_1}{\partial Q_i}
\end{equation}
\begin{equation}
K=H+\frac{\partial F_1}{\partial t}
\end {equation}
If now we were to set 
\begin{equation}
F=F_2(q_i,P_i)-Q_iP_i
\end{equation}
then it follows in a similar fashion that
\begin{equation}
p_i=\frac{\partial F_2}{\partial q_i} 
\end{equation}
\begin{equation}
Q_i=\frac{\partial F_2}{\partial P_i}
\end{equation}
\begin{equation}
K=H+\frac{\partial F_2}{\partial t}.
\end{equation}
Next, if we take
\begin{equation}
F=F_3(p_i,Q_i)+q_ip_i
\end{equation}
we obtain
\begin{equation}
q_i=-\frac{\partial F_3}{\partial p_i}
\end{equation}
\begin{equation}
P_i=-\frac{\partial F_3}{\partial Q_i}
\end{equation}
\begin{equation}
K=H+\frac{\partial F_3}{\partial t}.
\end{equation}
Lastly, there is the case in which
\begin{equation}
F=F_4+q_ip_i-Q_iP_i;
\end{equation}
this results in
\begin{equation}
q_i=-\frac{\partial F_4}{\partial p_i}
\end{equation}
\begin{equation}
Q_i=\frac{\partial F_4}{\partial P_i}
\end{equation}
\begin{equation}
K=H+\frac{\partial F_4}{\partial t}.
\end{equation}

    The transformation generated by $F_2$ is the identity if 
\begin{equation}
F_2=q_iP_i;
\end{equation}    
an infinitesmal canonical transformation takes place if
\begin{equation}
F_2=q_iP_i+\epsilon G(q_i,P_i,t)
\end{equation}
where $\epsilon$ is a small parameter. In this case eqs. (39) and (40) result in 
\begin{equation}
\delta p_i\equiv P_i-p_i=-\epsilon \frac{\partial G}{\partial q_i}
\end{equation}
\begin{equation}
\delta q_i\equiv Q_i-q_i=\epsilon \frac{\partial G}{\partial P_i}. 
\end{equation}
Upon taking $\epsilon=\delta t$, a small increment in time, and identifying $G$ with the Hamiltonian $H$,
these equations are seen to be identical to Hamilton's equations, so that the time evolution of a system can 
be viewed as  sequence of canonical transformations. Similarly, a translation in a direction $i$ by an amount $\epsilon$ is generated by the momentum $p_i$ (that is, we just take $G=p_i\approx P_i$), and a rotation about an axis i through an angle $\epsilon$ is generated by the angular momentum 
\begin{equation}
L_i=\epsilon_{ijk}q_jp_k.
\end{equation}
The fundamental Poisson brackets lead to the interesting result that
\begin{equation}
\left\{L_i,L_j\right\}=\epsilon_{ijk}L_k.
\end{equation}

    The notion of a canonical transformation can also be used to reduce a dynamical problem to that of solving a single partial differential equation, the Hamilton-Jacobi equation. The Hamilton-Jacobi function is a transformation function $F_2(q_i,P_i,t)$, often denoted by $S$, that results in a vanishing new Hamiltonian $K$. Eq. (41) then reduces to
\begin{equation}
0=H\left(q_i,\frac{\partial S}{\partial q_i}\right)+\frac{\partial S}{\partial t}.
\end{equation}
In this equation, we have used eq. (39) to eliminate $p_i$ in the argument of $H(q_i,p_i)$ in favour of $\frac{\partial S}{\partial q_i}$. 

    The momenta $P_i$ are necessarily time independent as the new Hamitonian $K$ vanishes; these momenta are identified as being the constants of integration that arise in the course of solving for $S$ in eq. (56).
Calling these constants $\alpha_i$, we see that eq. (40) implies an additional set of constants given by
\begin{equation}
\beta_i=\frac{\partial S}{\partial \alpha_i}.
\end{equation}
Once we have solved eq. (56) for $S$ we can then use eq. (57) to obtain
\begin{equation}
q_i=q_i(\alpha_i,\beta_i,t).
\end{equation}
Thus the constants $\alpha$ and $\beta$ can be identified with the boundary conditions on the initial configuration of the system that are needed in the course of solving for its dynamical evolution. 

     A formal result is that 
\begin{equation}
\frac{dS(q_i(t),\alpha_i,t)}{dt}=\frac{\partial S}{\partial q_i}\dot{q}_i+\frac{\partial S}{\partial t}
\end{equation}
which by eqs. (40) and (56) yields
\begin{equation}
S=\int (p_i\dot{q}_i-H)dt.
\end{equation}
This equation shows that the action $S$ appearing in eq. (20) is in fact identical to the Hamilton-Jacobi function. This doesn't make it possible to find the Hamilton-Jacobi function though, as the integral of eq. (60) can only be evaluated once $q_i(t)$ and $p_i(t)$ are known.

     As a simple example, let us consider a one dimensional harmonic oscillator in which both the mass and angular frequency are scaled to one. In this case, the kinetic and potential energies are given by
\begin{equation}
T=\frac{1}{2}\dot{q}^2    
\end{equation}     
\begin{equation}
V=\frac{1}{2}q^2
\end{equation}
so that with $L=T-V$ and $H$ given by eq. (13), we have
\begin{equation}
H=\frac{1}{2}(p^2+q^2).
\end{equation}
If the Hamilton-Jacobi function $S$ is taken to have the form
\begin{equation}
S(q,\alpha ,t)=W(q,\alpha )-\alpha t
\end{equation}
then eq. (56) in this case reduces to
\begin{equation}
\frac{1}{2}\left[\left(\frac{\partial W}{\partial q}\right)^2+q^2\right]=\alpha. 
\end{equation}
(The constant $\alpha$ is evidently the energy $E$ of the oscillator.) As a result of this equation, we find that 
\begin{equation} 
S=\int dq \sqrt{2\alpha-q^2}-\alpha t.
\end{equation}
Now using eq. (57), we arrive at
\begin{equation}
\beta =\int \frac{dq}{\sqrt{2\alpha-q^2}} -t
\end{equation}
which can be integrated to yield
\begin{equation}
q=\sqrt{2\alpha}sin(t+\beta ).
\end{equation}

    Hamilton himself recognized that in the case where 
\begin{equation}
H=\frac{1}{2m}p^2+V(q)
\end{equation}
the Hamilton-Jacobi equation is related to the wave equation.
If a wave travels in empty space with velocity $c$ and is in a medium with refractive index $\mu (q)$ , then the wave equation is 
\begin{equation}
\nabla ^2\psi -\frac{\mu ^2}{c^2}\frac{\partial^2\psi}{\partial t^2} =0.
\end{equation} 
Restricting our attention to a single frequency $\omega=kc$, we examine solutions to eq. (70) that are of the form 
\begin{equation}
\psi (q,t)=\psi_0 (q)e^{i(kf(q)-\omega t)}.
\end{equation}
For large values of $k$ (in other words, short wave lengths), eq. (70) then reduces to
\begin{equation}
(\nabla f)^2=\mu^2.
\end{equation}
If one were to take the form of the Hamilton-Jacobi function $S$ associated with the Hamiltonian of eq. (69) to be 
\begin{equation}
S=akf(q)-Et
\end{equation}
then the Hamilton-Jacobi eq. (56) with this Hamiltonian becomes 
\begin{equation}
\frac{1}{2m}(ak\nabla f(q))^2+V(q)=E
\end{equation}
which is the same as eq. (72) provided 
\begin{equation}
\mu^2=\frac{2m(E-V)}{a^2k^2}.
\end{equation}
This identification means that the original wave eq. (70) would become 
\begin{equation}
\nabla^2\psi +\frac{2m(E-V)}{a^2}\psi=0.
\end{equation}
The constant $a$ can now be identified with $\hbar$ so that eq. (76) becomes the time independent Schrodinger equation.

\subsection*{1.4 The Dirac Constraint Formalism and Gauge Invariance}

    In some cases it turns out that it is convenient to consider Lagrangians in which not all of the variables in configuration space are independent. (This may be to ensure that symmetries present in the system being considered are readily apparent.) In this case the equations of motion do not uniquely determine the evolution of the system as time elapses. The Dirac constraint formalism provides a systematic way of treating such systems.

    To appreciate the sort of difficulties that can arise in extracting information about the time evolution of such systems, consider a Lagrangian $L(q_i(t),\dot{q}_i(t))$  $(i=1...N)$ with $q_i(0)$ and $\dot{q}_i(0)$ being prescribed initial conditions. If we now examine a Taylor series expansion of $q_i(t)$ about $t=0$, then
\begin{equation}
q_i(t)=q_i(0)+t\dot{q}_i(0)+\frac{1}{2}t^2\ddot{q}_i(0)+.....
\end{equation}
with the boundary conditions providing the first two terms of this expansion. In the third term, the value of $\ddot{q}_i(0)$ is in principle fixed in terms of $q_i(0)$ and $\dot{q}_i(0)$ by Lagrange's eq. (3). This is because eq. (3) can be expessed in the form
\begin{equation}
\frac{\partial^2L}{\partial \dot{q}_i\partial \dot{q}_j}\ddot{q}_j+\frac{\partial^2L}{\partial \dot{q}_i\partial q_j}\dot{q}_j-\frac{\partial L}{\partial q_i}=0,
\end{equation}
allowing us to solve for $\ddot{q}_i(0)$ provided the so-called Hessian matrix 
\begin{equation}
\texttt{M}_{i,j}=\frac{\partial^2L}{\partial \dot{q}_i\partial \dot{q}_j}
\end{equation}
can be inverted. (Taking the time derivative of the Lagrange equation allows us in principle to find higher derivatives of $q_i$ at time $t=0$.) Systems in which this inversion is not possible contain constraints and do not have the values of $\ddot{q}_i(0)$ (and consequently of $q_i(t)$) fixed in terms of the boundary conditions.

    Inability to invert the Hessian matrix amounts to being unable to solve eq. (12) so that the generalized velocities $\dot{q}_i$ are expressed in terms of the canonical momenta $p_i$ without the imposition of a set of ``primary'' constraints 
\begin{equation}
\pi_i(q,p)=0.
\end{equation}    
The number of these primary constraints is $N-R$ where $R$ is the rank of the Hessian matrix.

    Once these constrants are satisfied, it becomes possible to define the canonical Hamiltonian $H_C$ as in eq. (13); however, in order 
    that eq. (80) is satisfied we consider the ``total'' Hamiltonian
\begin{equation}
H_T=H_C+u_i\pi_i
\end{equation}
where the fields $u_i$ are a set of Lagrange multiplier fields. The equations of motion for these fields ensure that the primary constraints are satified.

    However, for consistency it is also necessary to have 
\begin{equation}
\frac{d\pi_i}{dt}\approx 0
\end {equation} where $\approx$ denotes a ``weak'' equality, one which holds when the constraints themselves are satisfied. Using eq. (23), this means that 
\begin{equation}
\left\{\pi_i,H_T\right\}\approx 0.
\end{equation}
This condition may be automaticly satisfied. It may also be solved for some or all of the Lagrange multiplier $u_i$ in eq. (81). Lastly, it may require 
imposition of additional constraints
\begin{equation}
\sigma_i (q,p)=0,
\end{equation}
where these $\sigma_i$ are known as ``secondary'' constraints. The process applied to the primary constraints is now repeated; we consider 
\begin{equation}
\left\lbrace \sigma_i, H_T \right\rbrace \approx 0 
\end{equation}
and check to see if the time derivative of all of constraints, both primary and secondary, vanishes weakly. Further tercery constraints may arise (though this does not happen often). One continues adding constraints until all constraints have a vanishing Poisson bracket with the Hamiltonian when the constraints themselves are satisfied.

    Having obtained all constraints ($\psi_i$), it is convenient to divide them into first class constraints ($\phi_i$) and second class constraints ($\theta_i$). A first class constraint has a weakly vanishing Poisson bracket with all other constraints so that
\begin{equation}
\left\{\phi_i,\psi_j\right\}\approx 0;
\end{equation}
a second class constraint is one that is not first class. (Separating the first class constraints from the second class constraints can be difficult.) This distinction between the two classes of constraints is important, as the first class constraints are associated with arbitrariness in the time development of the system while second class constraints can be used to eliminate degrees of freedom in phase space. To see this, we consider the extended Hamiltonian $H_E$ in the form
\begin{equation}
H_E=H_C+c_i\phi_i+d_i\theta_i
\end{equation}
with the Lagrange multipliers $c$ and $d$ associated with the first and second class constraints respectively. The requirement 
\begin{equation}
\left\{\phi_i,H_E\right\}\approx 0
\end{equation}
is satisfed for all values of $c_i$ and $d_i$ as first class constraints have a weakly vanishing Poisson bracket with both first and second class constraints. However, when we consider
\begin{equation}
\left\{\theta_i,H_E\right\}\approx 0
\end{equation}
we find a set of conditions on the Lagrange multipliers $d_i$; otherwise $\theta_i$ would be first class. 
If now we apply eq. (23) to examine how a dynamical variable $A$ evolves in time, we find that since the Lagrange multipliers $c_i$ appearing in $H_E$ are arbitrary, $\frac{dA}{dt}$ is not fixed.

    Elimination of superfluous degees of freedom using second class constraints involves use of Dirac brackets in place of Poisson brackets. This entails first defining the antisymmetric matrix
\begin{equation}
d_{i,j}\approx \left\{\theta_i,\theta_j\right\}.
\end{equation}    
This matrix is invertible; otherwise not all of the constraints $\theta_i$ would be second class. The number of second class constraints is consequently even because the dimension of any invertible antisymmetric matrix must be even. We now define the Dirac bracket of two dynamical variables $A$ and $B$ to be
\begin{equation}
\left\{A,B\right\}^*\equiv\left\{A,B\right\}-\left\{A,\theta_i\right\}d_{i,j}^{-1}\left\{\theta_j,B\right\}.
\end{equation}
These brackets have a number of useful properties. First of all, we note that for any second class constraint $\theta_i$ we have
\begin{equation}
\left\{\theta_i,B\right\}^*=0.
\end{equation}
Next, if $F$ is a first class dynamical variable (that is, one whose Poisson bracket with any constraint $\psi_i$ weakly vanishes so that $\left\{F,\psi_i\right\}\approx 0$), then it follows immediately that 
\begin{equation}
\left\{F,B\right\}\approx \left\{F,B\right\}^*.
\end{equation}
The Hamiltonian is such a first class quantity, as can be seen from eq. (23) and the consistency condition that all constraints have a vanishing time derivative.
As a result of eq. (93) then, we can replace eq. (23) by
\begin{equation}
\frac{dA}{dt}\approx \left\{A,H_T\right\}^*;
\end{equation}
this equation need hold only weakly, that is, on the constraint surface. But now eq. (92) tells us that in $H_T$ we can set the second class constraints $\theta_i$ equal to zero. It is consequently not necessary to determine the value of the Lagrange multipliers $d_i$ appearing in eq. (87). Thus by use of Dirac brackets, we have been able to eliminate some of the degrees of freedom in phase space through setting the second class constraints equal to zero. 

    The Dirac brackets can be shown to satisfy eqs. (26-29); proving the Jacobi identity  (29) for the Dirac bracket is a tedious exercise.

    It is still necessary to deal with the arbitrariness noted above that would follow from the possible presence of first class constraints. Rather than simply choosing a value for the Lagrange multipliers $c_i$ appearing in eq. (87), we introduce extra conditions, known as gauge conditions, in phase space
\begin{equation}
\gamma_i(q_i,p_i)=0
\end{equation}
with one gauge condition for each first class constraint. (The reason for calling these extra conditions ``gauge conditions'' becomes apparent later when the Dirac constraint approach is used to analyze the canonical structure of the electromagnetic field.) These gauge conditions are arbitrary functions of $q_i$ and $p_i$; they are independent of $t$ and the velocities $\dot{q}_i$ and $\dot{p}_i$. We also require that these gauge conditions, together with the first class constraints $\phi_i$, form a set of second class constraints. In other words, we impose the restriction that if 
\begin{equation}
\texttt{D}_{i,j}\approx \left\{\phi_i,\gamma_j\right\}
\end{equation}
then 
\begin{equation}
det\texttt{D}\neq 0.
\end{equation}
As a result, the first class constraints, the second class constraints and the gauge conditions together form a set of constraints which can be treated as the second class constraints were by themselves. If $\Theta_i$ denotes the set $\left\{\phi_i,\theta_i,\gamma_i\right\}$, then Dirac brackets can be defined as in eq. (91), with $\Theta_i$ replacing $\theta_i$.

    We now examine a simple system that serves to illustrate how Dirac brackets can be used. Consider a system in which a particle is constrained to move in a circle of unit radius with constant speed. Obviously, the easiest approach would be to use the angle $\theta (t)$ which gives the position of the particle as the only variable in phase space. Instead though, suppose we perversely use the Cartesian coordinates $x(t)$ and $y(t)$ to describe the position of the particle and in this two dimensional configuration space consider the Lagrangian
\begin{equation}
L=\frac{1}{2}(\dot{x}^2+\dot{y}^2-x^2-y^2)+\lambda_1\left(\dot{x}-y\right)+\lambda_2\left(\dot{y}+x\right).
\end{equation}
The fields $\lambda_1$ and $\lambda_2$ are Lagrange multiplier fields whose equations of motion ensure that the particle moves in a circular path at constant speed; their presence means that configuration space is now four dimensional.

    The momenta associated with $x$, $y$, $\lambda_1$ and $\lambda_2$ are 
\begin{equation}
p_x=\dot{x}+\lambda_1    
\end{equation}
\begin{equation}
p_y=\dot{y}+\lambda_2
\end{equation}
\begin{equation}
p_{\lambda_1}=0=p_{\lambda_2}.
\end{equation}
The last equation constitutes a pair of primary constraints as one cannot solve for either $\dot{\lambda_1}$
or $\dot{\lambda_2}$ in terms of $p_{\lambda_1}$ and $p_{\lambda_2}$.

    However, if we use these primary constraints, we can define the canonical Hamiltonian
\begin{equation}
H_C=p_x\dot{x}+p_y\dot{y}+p_{\lambda_1}\dot{\lambda_1}+p_{\lambda_2}\dot{\lambda_2}-L
\end{equation}
and find that with the Lagrangian of eq. (94)
\begin{equation}
H_C=\frac{1}{2}\left[\left(p_x-\lambda_1\right)^2+\left(p_y-\lambda_2\right)^2+x^2+y^2\right]+\lambda_1y-\lambda_2x.
\end{equation}
 
    Applying eq. (83) with the primary constraints of eq. (101) leads to a pair of secondary constraints
\begin{equation}
p_x-\lambda_1-y=0=p_y-\lambda_2+x.     
\end{equation}
Together, eqs. (83) and (104) make up a set of four second class constraints, as $\left\{\lambda_i,p_{\lambda_j}\right\}=\delta_{i,j}$. It is quite easy to compute the matrix $d_{i,j}$ of eq. (90) associated with these four constraints; once that is done, it is possible to define the Dirac bracket of eq. (91) and we finally end up with the fundamental Dirac brackets
\begin{equation}
\left\{x,p_x\right\}^*=1=\left\{y,p_y\right\}^*
\end{equation}
with all other fundamental Dirac brackets vanishing. The Hamiltonian of eq. (103) itself, upon treating the constraints as strong equations, reduces to simply 
\begin{equation}
H_C=p_xy-p_yx.
\end{equation}   
With this Hamiltonian, we find that
\begin{equation}
\dot{x}\approx \left\{x,H\right\}^*=y
\end{equation}
and
\begin{equation}
\dot{y}\approx \left\{y,H\right\}^*=-x
\end{equation}
as anticipated.

    The simple system just considered does not contain any first class constraints. Before dealing with a particular situation in which first class constraints do occur, let us examine the nature of the gauge transformations arising due to the presence of first class constraints. Suppose we restrict ourselves to the situation in which there are only first class constraints, and these are either primary $(\pi 1_i)$ or secondary $(\pi 2_i)$, with $\pi 2_i$ arising from the requirement that $\frac{d\pi 1_i}{dt}$ vanish when the primary constraints are zero. Inclusion of  tercery constraints and beyond, as well as second class constraints, is quite easy. 
    
    The arbitrariness associated with the presence of first class constraints means that trajectories in phase space are not unique; trajectories given by $q_i\left(t\right)$ and $p_i\left(t\right)$ as well as by $q_i(t)+\alpha_i(t)$ and $p_i(t)+\beta_i(t)$ can both be physical. We now take the generator that relates these two trajectories to be a function $G$ whose form is given by
\begin{equation}
G=\epsilon(t)G_0+\dot{\epsilon}(t)G_1    
\end{equation}
where $\epsilon$ is an infinitesmal parameter that may contain time dependence. (If there were any tercery constraints present, then a term dependent on $\ddot{\epsilon}(t)$ would also contribute to $G$.) As this function $G$ is a generator, we have the equations
\begin{equation}
\alpha_i(t)=\left\{q_i,G\right\}=\epsilon (t)\frac{\partial G_0}{\partial p_i}+\dot{\epsilon}(t)\frac{\partial G_1}{\partial p_i}
\end{equation}
and
\begin{equation}
\beta_i(t)=\left\{p_i,G\right\}=-\epsilon (t)\frac{\partial G_0}{\partial q_i}-\dot{\epsilon}(t)\frac{\partial G_1}{\partial q_i}.
\end{equation}
From eq. (110) we see that 
\begin{equation}
\dot{\alpha}_i\approx \dot{\epsilon}\frac{\partial G_0}{\partial p_i}+\epsilon \left\{\frac{\partial G_0}{\partial p_i},H_T\right\}+\ddot{\epsilon}\frac{\partial G_1}{\partial p_i}+\dot{\epsilon}\left\{\frac{\partial G_1}{\partial p_i},H_T\right\}.
\end{equation}
(We take the Hamitonian $H_T$ to only incorporate the primary first class constraints in this discussion; weak inequalities hold on the constraint surface $\pi 1_i=0$ and not necessarly when $\pi 2_i=0$.) The weak version of eq. (16) also holds for both $q_i$ and $q_i+\alpha_i$; we hence find that 
\begin{equation}
\dot{q}_i+\dot{\alpha_i}\approx \frac{\partial H_T(q_i+\alpha_i,p_i+\beta_i)}{\partial p_i}.
\end{equation}
To lowest order in $\alpha_i$ and $\beta_i$, eq. (113) leads to
\begin{equation}
\dot{\alpha}_i\approx \alpha_j\frac{\partial^2H_T}{\partial p_i\partial q_j}+\beta_j\frac{\partial^2H_T}{\partial p_i\partial p_j}.
\end{equation}
We can equate $\dot{\alpha}_i$ found from eqs. (110) and (114) and then use eqs. (110) and (111) to eliminte $\alpha_i$ and $\beta_i$ to obtain
\begin{equation}
\frac{\partial}{\partial p_i}\left[\epsilon \left\{G_0,H_T\right\}+\dot{\epsilon}(G_0+\left\{G_1,H_T\right\})+\ddot{\epsilon}G_1\right]\approx 0.
\end{equation}
Working with the equation for $\dot{\beta}_i$ that is analogous to eq. (114), it is also possible to find that
\begin{equation}
\frac{\partial}{\partial q_i}\left[\epsilon 
\left\{G_0,H_T\right\}+\dot{\epsilon}(G_0+\left\{G_1,H_T\right\})+\ddot{\epsilon}G_1\right]\approx 0.
\end{equation}
Together, eqs. (115) and(116) can be solved by taking
\begin{equation}
G_1=\pi 1_i,
\end{equation}
and then if $\pi 2_i$ is found by considering $\left\lbrace \pi 1_i, H_T\right\rbrace$, 
\begin{equation}
 G_0=\left(-\pi 2_i+\lambda_{ij}\pi 1_j\right)
\end{equation}
where $\lambda_{ij}$ is finally determined by the requirement that
\begin{equation}
\left\{G_0,H_T\right\}\approx 0.
\end{equation}
We see that there is a separate generator $G$ and associated gauge function $\epsilon$ for each of 
the first class primary constraints $\pi 1_i$.  
This way of finding the generators of gauge transformations will prove to be quite useful when we come to analyze field theories containing first class constraints.

An alternate approach to finding the gauge invariance of the classical action involves considering the ``extended action''
\begin{equation}
S_E = \int_{t_{1}}^{t_{2}} dt\left( p_i \dot{q}_i - H_E\right)
\end{equation}
with
\begin{equation}
H_E = H_C + U^{a{_{i}}} \phi_{a{_{i}}} + U^{\alpha{_{i}}} \theta_{\alpha{_{i}}}
\end{equation}
where $\phi_{a{_{i}}}$ is an i$^{th}$ generation first class constraint and $\theta_{\alpha{_{i}}}$ is an i$^{th}$ generation second class 
constraint.  We know that
\begin{equation}
\left\lbrace \phi_a, \phi_b\right\rbrace = C_{ab}^{\;c} \phi_c + C_{ab}^{\alpha\beta} \theta_\alpha \theta_\beta + C_{ab}^\alpha \theta_\alpha
\end{equation}
\begin{equation}
\left\lbrace \phi_a, \theta_\alpha\right\rbrace = C_{a\alpha}^{\;b} \phi_b + C_{a\alpha}^{\beta} \theta_\beta
\end{equation}
and that 
\begin{equation}
\left\lbrace H_C, \phi_a\right\rbrace = V_a^{\;b} \phi_b + V_a^{\alpha\beta} \theta_\alpha\theta_\beta + V_a^\alpha \theta_\alpha
\end{equation}
\begin{equation}
\left\lbrace H_C, \theta_\alpha\right\rbrace = V_\alpha^{b} \phi_b + V_\alpha^\beta \theta_\beta\,.
\end{equation}
(The form of $C_{ab}^\alpha$ and $V_a^\beta$ is constrained as $\left\lbrace \phi_a, \phi_b\right\rbrace$  and $\left\lbrace H_C, \gamma_a\right\rbrace$ must be 
first class.)  If now $q_i$ and $p_i$ undergo the ``gauge transformations''
\begin{equation}
\delta_\epsilon q_i = \epsilon^a(t) \frac{\partial\phi_a}{\partial p_i}\qquad \delta_\epsilon p_i = -\epsilon^a(t) \frac{\partial\phi_a}{\partial q_i}
\end{equation}
while
\begin{equation}
\delta_\epsilon U^a = \dot{\epsilon}^a + U^c\epsilon^b C_{bc}^a + U^\alpha\epsilon^b C^a_{b\alpha} - \epsilon^b V_b^a
\end{equation}
\begin{equation}
\delta_\epsilon U^\alpha = U^c\epsilon^b \left(C_{bc}^\alpha + C_{bc}^{\alpha\beta} \theta_\beta\right) - \epsilon^b \left(V_b^\alpha
 + V_b^{\alpha\beta} \theta_\beta\right)\\
+ U^\beta\epsilon^b C_{b\beta}^\alpha
 \end{equation}
then
\begin{equation}
\delta_\epsilon \left(p_i \dot{q}_i - H_E\right) = \frac{d}{dt} \left(-\epsilon^a\phi_a + p_i\epsilon^a \frac{\partial\phi_a}{\partial p_i}\right).
\end{equation}
Consequently, if $\epsilon^a(t_1) = \epsilon^a(t_2) = 0$, the action of eq. (120) is unaltered by a gauge transformation.

Eqs. (126-128) can be recast into the form
\begin{equation}
\overline{\delta} F = \left\lbrace F, \mu^a \phi_a\right\rbrace 
\end{equation}
where $\mu^a$ now depends not only on $t$ but also $p_i$, $q_i$, $U^a$ and $U^\alpha$. In this case
\begin{equation}
\overline{\delta} U^a = \frac{D\mu^a}{Dt} + \left\lbrace \mu^a, H_E\right\rbrace + U^c \mu^b C_{bc}^a\\
+ U^\alpha\mu^b C_{b\alpha}^a - \mu^b V_b^a
\end{equation}
\begin{equation}
\overline{\delta} U^\alpha = U^c \mu^b \left( C_{bc}^\alpha + C_{bc}^{\alpha\beta} \theta_\beta\right) - \mu^b \left(V_b^\alpha + V_b^{\alpha\beta}\theta_\beta\right)\\
+ U^\beta\mu^b C_{b\beta}^\alpha \,,
\end{equation}
where
\begin{equation} \frac{D}{Dt} = \frac{\partial}{\partial t} + \left( \dot{U}^a \frac{\partial}{\partial U^a} + \ddot{U}^a \frac{\partial}{\partial \dot{U}^a} + \ldots\right)\\
+ \left(\dot{U}^\alpha \frac{\partial}{\partial U^\alpha} + \ddot{U}^\alpha \frac{\partial}{\partial \dot{U}^\alpha} + \ldots \right)
\end{equation}
is the total time derivative, exclusive of dependency on time through $q_i$ and $p_i$.  It then follows that
\begin{equation}
\overline{\delta} \left(p_i\dot{q}_i - H_E\right) = \frac{d}{dt} \left[-\mu^a\phi_a + p_i \frac{\partial}{\partial p_i} (\mu^a\phi_a)\right]\,.
\end{equation}

The transformation that leaves the action of eq. (4) invariant can be deduced by considering the invariance of
\begin{equation}
S_T = \int_{t{_1}}^{t{_2}} dt \left(p_i \dot{q}_i - H_C - U^{a{_1}} \phi_{a{_1}} - U^{\alpha{_1}} \theta_{\alpha{_1}}\right)
\end{equation}
where only primary constraints occur in the sum appearing in $S_T$.  As $S_E$ of eq. (120) reduces to $S_T$ of eq. (135) upon setting $U^{a{_2}} \ldots U^{a{_m}}$, 
$U^{\alpha{_2}} \ldots U^{\alpha{_n}}$ equal to zero, the invariance of $S_E$ is determined by choosing the ``gauge'' in which $U^{a{_2}} = \ldots = U^{\alpha{_n}} = 0 = \delta U^{a_{2}} \ldots = \delta U^{\alpha_{n}}$ 
and iteratively solving for $\mu^{a{_1}} \ldots \mu^{a{_m}}$ using eqs. (131,132).  The number of gauge functions $\mu^a$ then equals the number of primary first class 
constraints and their time derivatives occur if there are secondary, tercery etc. first class constraints.

The procedure used to fix $\mu^{a_{i}}$ involves equating the coefficients of the constraints arising in eq. (131) to zero after setting $U^{\alpha_{i}}$, $U^{a_{i}}$, $\delta U^{\alpha_{i}}$, $\delta U^{a_{i}}\;(i = 2 \ldots N)$ to zero.  If $ i = 1 \ldots N$ (i.e. there are $N$ generations of constraints) then we first take $\mu^{a_{N}}$ to be solely dependent on $t$, and then determine $\mu^{a_{1}} \ldots \mu^{a_{N-1}}$ are found in terms of $\mu^{a_{N}}$.  These coefficients are uniquely determined by this procedure provided there is the same number of constraints occurring in each generation.  This does not always happen; in a model in which a scalar field is on a curved surface in $1 + 1$ dimensions there are two generations of constraints with more primary than secondary constraints and so the coefficients $\mu^{a_{1}}$ are not fixed by $\mu^{a_{2}}$.

\subsection*{\large 1.5 Grassmann Variables and their Canonical Formalism}

    Grassmann variables have the distinctive property that they anticommute under multiplication. That is, if $\theta_1$ and $\theta_2$ are Grassmann variables, then
\begin{equation}
\theta_1\theta_2=-\theta_2\theta_1
\end{equation}
As a result, we cannot ``count'' with Grassmann variables, and they are not ordered; no meaning can be given to an inequality such as $\theta_1>\theta_2$. (For Hermitian conjugation we have $(\theta_1\theta_2)^\dagger = \theta_2^\dagger \theta_1^\dagger . $)

    Any Taylor series expansion of a function of Grassmann variables in powers of these variables must terminate after a finite number of terms since if $\theta$ is Grassmann, then $\theta^2=0$ on account of eq. (136). Thus a function $F(x;\theta_1,\theta_2,...\theta_n)$ can be expanded 
\begin{equation}
F(x;\theta_1,...,\theta_n)=f_0(x)+f_1(x)_i\theta_i+f_2(x)_{i_1i_2}\theta_{i_1}\theta_{i_2}+...+f_n(x)_{i_1i_2...i_n}\theta_{i_1}\theta_{i_2}...\theta_{i_n}
\end{equation}
where the functions $f_m(x)_{i_1...i_m}$ are all antisymmetric in the indices $i_1...i_m$.

    One can introduce a calculus of Grassmann variables. Differentiation is defined so that 
\begin{equation}
\frac{d}{d\theta_i}\theta_j=\delta_{ij}.
\end{equation}
It is also necessary when taking derivatives to take into account the fact that Grassmann variables anticommute, so for example $\frac{d}{d\theta_2}\theta_1\theta_2=-\theta_1$ and $\frac{d^2}{d\theta_1d\theta_2}F=-\frac{d^2}{d\theta_2d\theta_1}F$. We also have $\frac{d}{dt} F(\theta (t)) = \dot{\theta}(t) F^\prime (\theta (t))$.
 
    Integration is defined to be the same as differentiation; we take
\begin{equation}
\int d^n\theta F\equiv \int d\theta_1d\theta_2...d\theta_nF\equiv \frac{d^n}{d\theta_1d\theta_2...d\theta_n}F
\end{equation}
so that, for example, $\int d\theta_1d\theta_2(\theta_1\theta_2)=-1$. It also follows that a shift of variables leaves a Grassmann integral unaltered so that $\int d\theta \theta=\int d\theta (\theta +\phi)$ where $\phi$ is also Grassmann.

    Under a change of variables $\phi_i = a_{ij}\theta_j$, it is evident from eq. (139) that 
\begin{equation}
\int d^n\theta F(\theta )= \int d^n\phi det(a_{ij}) F(\phi )
\end{equation}    
and hence $d^n\theta = d^n\phi det(a_{ij})$. If $\theta$ and $\phi$ were ordinary variables rather than Grassmann, then 
the usual rules of calculus give $d^n\theta =d^n\phi det^{-1}(a_{ij})$. 

    If \textbf{x} is a real vector, then the standard Gaussian integral $\int^{\infty}_{-\infty} d\lambda e^{-a\lambda^2}=\sqrt{\frac{\pi}{a}}$ easily leads to
\begin{equation}
\int^{\infty}_{-\infty}d^n\textbf{x}exp\left(-\textbf{x}^T\textsl{A}\textbf{x}\right)=\frac{\pi^{\frac{n}{2}}}{\sqrt{det\textsl{A}}}
\end{equation}
for a (symmetric) matrix \textsl{A}. We also find that if \textbf{z} is a complex vector and \textsl{H} is an Hermitian
matrix, then
\begin{equation}
\int^{\infty}_{-\infty}d^n\textbf{z}d^n\textbf{z}^*exp\left(-\textbf{z}^{T*}\textsl{H}\textbf{z}\right)=\frac{\pi^n}{detH}.
\end{equation}

    It is also possible to deal with Gaussian integrals over Grassmann variables. If $\theta$ is an $n$ component Grassmann vector with $n$ even, and $\textsl{B}$ an $n$ by $n$ antisymmetric matrix, then eq. (139) can be used to show that 
\begin{equation}
\int d^n\theta exp\left(-\frac{1}{2}\theta^T \textsl{B}\theta \right)=\sqrt{det\textsl{B}}.
\end{equation}
Proving this involves making a transformation on $\textsl{B}$ so that it has a sequence of 2 by 2 blocks along its diagonal with zeros everywhere else. Furthermore, we can show that if $\theta$ and $\phi$ are independent Grassmann vectors of dimension $n$, then
\begin{equation}
\int d^n\phi d^n\theta exp\left(-\phi^T\textsl{H}\theta \right)=detH
\end{equation}
for a matrix $H$.

    There is also a canonical formalism associated with Grassmann quantities that are dynamical variables. A Lagrangian in this case is a function of both ordinary variables $q_i(t)$ and Grassmann variables $\theta_i (t)$. Variation of the action
\begin{equation}
S=\int^{t_2}_{t1} dtL(q_i(t),\dot{q}_i(t);\theta_i (t),\dot{\theta}_i(t))
\end{equation}
leads to the expected equations; eq. (3) as well as analogous equation for $\theta_i$. If we now define a Grassmann momentum conjugate to $\theta_i$ ,
\begin{equation}
\pi_i=\frac{\partial L}{\partial \dot{\theta}_i}
\end{equation}
as well as a Hamiltonian $H$
\begin{equation}
H(q_i,p_i;\theta_i,\pi_i)=\dot{q}_ip_i+\dot{\theta}_i\pi_i-L
\end{equation}
then variation of the action 
\begin{equation}
S=\int^{t_2}_{t_1} dt\left(\dot{q}_ip_i+\dot{\theta}_i\pi_i-H\right)
\end{equation}
leads to the canonical equations of motion of eqs. (16) and (17) as well as the equations
\begin{equation}
\dot{\theta}_i=-\frac{\partial H}{\partial \pi_i}
\end{equation}
and
\begin{equation}
\dot{\pi}_i=-\frac{\partial H}{\partial \theta_i}.
\end{equation}
The unexpected minus sign in eq. (149) comes about as $\dot{\theta}_i\delta \pi_i =-\delta \pi_i \dot{\theta}_i$.

    In order to retain eq. (23), we define Poisson brackets to be now given by
\begin{equation}    
\left\{A,B\right\}=\left(\frac{\partial B}{\partial p_i}\frac{\partial A}{\partial q_i}-\frac{\partial B}{\partial q_i}\frac{\partial A}{\partial p_i}\right)-\left(\frac{\partial B}{\partial \theta_i}\frac{\partial A}{\partial \pi_i}+\frac{\partial B}{\partial \pi_i}\frac{\partial A}{\partial \theta_i}\right).
\end{equation} 
The order in which terms appear in eq. (151) is important, as $A$ and $B$ may be Grassmann, as well as $\theta$ and $\pi$. The ordering of Grassmann quantities can also lead to extra minus signs appearing in eqs. (26-29) when considering these generalized Poisson brackets.

\subsection*{\large 1.6 Special Relativity}

   Any physical event is taken to occur at a specific time and place, as described by Cartesian coordinates $(t,x,y,z)$. These coordinates are those used by a specific observer; some other observer moving with relative velocity \textbf{v} would use different coordinates which we take to be $(t',x',y',z')$. If the two observers are coincident at $t=t'=0$ with their axes are aligned at that time, and \textbf{v} is in the direction of the $x$ and $x'$ axes, then classically
we have the so-called Galilean transformation
\begin{equation}
x'=x-vt,
\end{equation}
\begin{equation}
y'=y,
\end{equation}
\begin{equation}
z'=z,
\end{equation}
and
\begin{equation}
t'=t
\end{equation}
where $v$ is the magnitude of \textbf{v}. This transformation is consistent with everyday experience; in particular, it implies that if $u=\frac{dx}{dt}$ and $u^{'}=\frac{dx^{'}}{dt^{'}}$, then 
\begin{equation}
u^{'}=u-v.
\end{equation}
However, the Maxwell equations which describe electricity and magnetism predict the existence of waves travelling with velocity $c$. If $c$ is to be universal constant, then this velocity must be the same according to all observers and eq. (156) breaks down. Einstein pointed out that this implies the relationship between the primed and unprimed coordinates used to label an event can no longer be given by eqs. (152-155); in particular the spatial and temporal coordinates must mix. (The actual transformation was discovered earlier by Voigt, Larmor and Lorentz, but they only viewed this as a useful invariance of the Maxwell equations.) 

    If the clocks and coordinate axes coincide in the primed and unprimed frames at $t=t^{'}=0$ and the relative velocity $v$ is again along the $x$ and $x^{'}$ axes, then we assume the linear transformation
\begin{equation} 
x^{'}=a_{11}x+a_{12}t,
\end{equation}
\begin{equation} 
y^{'}=y,
\end{equation}
\begin{equation} 
z^{'}=z,
\end{equation}
and
\begin{equation}
t^{'}=a_{21}x+a_{22}t.
\end{equation}

    First of all, the point $x^{'}=0$ coincides with $x=vt$ and so eq. (157) implies that 
\begin{equation}
a_{12}=-va_{11}.    
\end{equation}

    Since $c$ is to be a universal constant, a wave front emitted along the positive $x$ axis at time $t=0$ will be at $x=ct$ and $x^{'}=ct^{'}$ in the unprimed and primed frames respectively. Eqs. (157) and (160) together with eq. (161) then imply that
\begin{equation}
ct^{'}=a_{11}(c-v)t
\end{equation}
and
\begin{equation}
t^{'}=(a_{21}c+a_{22})t.
\end{equation}
So also, a wave front along the negative $x$ axis leads to
\begin{equation}
-ct^{'}=a_{11}(-c-v)t
\end{equation}
and
\begin{equation}
t^{'}=(-a_{21}c+a_{22})t.
\end{equation}
From eqs. (162-165) it follows that 
\begin{equation}
x^{'}=a_{11}(x-vt),
\end{equation}
\begin{equation}
t^{'}=a_{11}(-\frac{v}{c^2}x+t).
\end{equation}
But because of the symmetry between our two reference frames, we must also have
\begin{equation}
x=a_{11}(x^{'}+vt^{'})
\end{equation}
and 
\begin{equation}
t=a_{11}(\frac{v}{c^2}x^{'}+t^{'});
\end{equation}
together eqs. (166-169) lead to
\begin{equation}
a_{11}=\frac{1}{\sqrt{1-\frac{v^2}{c^2}}}\equiv \gamma.
\end{equation}
These transformations are generally called the ``boost transformations''. 

It immediately follows that if $\mathbf{u}=\frac{d\textbf{r}}{dt}$
 and $\mathbf{u}^\prime =\frac{d\textbf{r}^\prime}{dt^\prime}$      
\begin{equation}
u^{'}_x=\frac{u_x-v}{1-\frac{u_xv}{c^2}},
\end{equation}
\begin{equation}
u^{'}_y=\frac{u_y}{\gamma \left(1-\frac{u_xv}{c^2}\right)},
\end{equation}
\begin{equation}
u^{'}_z=\frac{u_z}{\gamma \left(1-\frac{u_xv}{c^2}\right)}.
\end{equation}
These transformtion equations are consistent with the speed $c$ being a universal constant. We shall henceforth scale $c$ to one.

    If two events occur at the same position $x$ and at times $t_1$ and $t_2$ in one frame, then according to eq. (166),
\begin{equation}
t^{'}_2-t^{'}_1=\gamma (t_2-t_1)>t_2-t_1;
\end{equation}
this is so-called ``time dilation''. Similarly, if two events occur at the same time $t^{'}$ but at two different positions $x^{'}_1$ and $x^{'}_2$ (with the same $y^{'}$ and $z^{'}$ coordinates), then by eq. (168) we have
\begin{equation}
x^{'}_2-x^{'}_1=\frac{x_2-x_1}{\gamma}<x_2-x_1
\end{equation}
which is known as ``length contraction''.

    These two phenomena lead to the so-called ``twin paradox''. Suppose a rocket travels from earth to a star a constant distance $D$ from earth then immediately returns, always moving with speed $v$. According to an observer on earth, the time taken for this trip is simply $\frac{2D}{v}$. However, because an observer on the rocket is moving with respect to the earth and the star, he sees the distance between the earth and star to be just $D\sqrt{1-v^2}$ on account of eq. (175), and so that according to this observer the time for the round trip is $\frac{2D\sqrt{1-v^2}}{v}$. But by eq. (166), since the departure and return of the rocket both take place at the same place when viewed from the earth-bound frame, this time in the rocket's frame corresponds to a time $\left(\frac{2D\sqrt{1-v^2}}{v}\right)\sqrt{1-v^2}=\left(1-v^2\right)\frac{2D}{v}\neq \frac{2D}{v}$ in the earth's frame. This inequality at first appears to be paradoxical; it can be accounted for though by noting that there is a sequence of events on earth during a time $2vD$ that are never simultaneous with either the out-bound observer (they are in his future) or the in-bound observer (they are in his past). 
    
    To see this in more detail, let us call the earth-bound, out-bound and in-bound observers $E$, $O$ and $I$ respectively, so that according to eq. (167)
\begin{equation}
t_O=\gamma (-vx_E+t_E)   
\end{equation}
and
\begin{equation}
t_I=\gamma (vx_E+t_E)   
\end{equation}
if the three frames used by these observers coincide at the instant the rocket initially leaves earth. Let us now consider three events; event $L$ is the landing of the rocket on the star and its subsequent take-off towards earth, event $A$ is an event on earth that is simultaneous \textsl{in the frame $O$} to $L$, and event $B$ is an event on earth that is simultaneous \textsl{in the frame $I$} to $L$. Event $L$ has coordinates $(t_E,x_E)=(D,\frac{D}{v})$. Furthermore, by eq. (176), event $L$ occurs at time $t^{L}_{O}=\sqrt{1-v^2}\frac{D}{v}$ according to the out-bound observer and by eq. (177) at time $t^{L}_{I}=\left(\frac{1+v^2}{\sqrt{1-v^2}}\right)\frac{D}{v}$ according to the in-bound observer. By the way in which event $A$ and $B$ are defined we know that $t^{A}_{O}=t^{L}_{O}$ and $t^{B}_{I}=t^{L}_{I}$. Finally, knowing that events $A$ and $B$ both occur at $x_E=0$, we see from eq. (176) that $t^{A}_{E}=\left(1-v^2\right)\frac{D}{v}$, and we see from eq. (177) that $t^{B}_{E}=\left(1+v^2\right)\frac{D}{v}$. Events occuring on earth between $t^{A}_{E}$ and $t^{B}_{E}$ are never simultaneous with either the out-going or in-coming observer on the rocket; there is consequently a sequence of events on earth during a time $t^{B}_{E}-t^{A}_{E}=2vD$; this precisely accounts for the inequality of the ``twin paradox'' noted above.

From eqs. (166-170) it follows that   
\begin{equation}      
t^{'2}-\textbf{r}^{'2}=t^2-\textbf{r}^2\;.
\end{equation}

    Eq. (178) is quite fundamental. We first note that it implies that the length of the four component vector 
$\textbf{x}=(it,\textbf{r})$ (where $\textbf{r}$ is itself the spatial vector $(x,y,z)$ and $it$ is an imaginary fourth component) is unaltered by the boost transformations of eqs. (166-170), where the length of $\textbf{x}$ is given by the usual dot product $\textbf{x}.\textbf{x}$. The boost transformation itself can then be taken to be a rotation in the $it-x$ plane, albeit through an imaginary angle $i\theta $ where 
\begin{equation}    
cos(i\theta )=\gamma ,\,\,sin(i\theta )=iv\gamma
\end{equation}
It is more convenient though to define a real four component vector $x^{\mu }=(t,\textbf{r})$ and to combine the boost transformation and spatial rotations into the so-called ``Lorentz'' transformation. (If we also include translations in both time and space, we have the ``Poincare'' transformations.) The Lorentz transformations leave 
\begin{equation}
x^2\equiv \eta_{\mu \nu}x^{\mu}x^{\nu}\equiv t^2-\textbf{r}^2
\end{equation}
unaltered. This four dimensional space with the magnitude of a vector being given by eq. (180) is called ``Minkowski space''. The tensor (two component matrix) $\eta_{\mu \nu}$ is diagonal with $\eta_{00}=-\eta _{11}=-\eta_ {22}=-\eta_{33}=1$; it is the Minkowski space metric. (Indices associated with time are generally given the label ``$0$''.) It is convenient to define $\eta^{\mu \nu}$ to be the inverse of $\eta_{\mu \nu}$ so that
\begin{equation}
\eta^{\mu \lambda}\eta_{\lambda \nu}=\delta^{\mu}_{\nu}.
\end{equation}

    A general Lorentz transformation is given by 
\begin{equation}
x^{'\mu}= \Lambda^{\mu}_{\;\;\;\nu}x^{\nu}.
\end{equation}
Eq. (180) places a restriction on the transformation matrix $\Lambda^{\mu}_{\;\;\;\nu}$,
\begin{equation}
\eta_{\mu \nu}=\eta_{\rho \sigma}\Lambda^{\rho}_{\;\;\;\mu}\Lambda^{\sigma}_{\;\;\;\nu}.
\end{equation}
If $\Lambda_{\mu}^{\;\;\;\nu}$ is defined to be the inverse of $\Lambda^{\mu}_{\;\;\;\nu}$ then it follows that
\begin{equation}
\Lambda^{\sigma}_{\;\;\;\mu}\Lambda_{\rho}^{\;\;\;\mu}=\delta^{\sigma}_{\rho}.
\end{equation} 
Eq. (182) now tells us that if $\partial_{\mu}\equiv \frac{\partial}{\partial x^{\mu}}$, then
\begin{equation}
\partial_{\mu}=\Lambda^{\nu}_{\;\;\;\mu}\partial^{'}_{\nu};
\end{equation}
eq. (184) converts eq. (185) to
\begin{equation}
\partial^{'}_{\mu}=\Lambda_{\mu}^{\;\;\;\nu}\partial_{\nu}.
\end{equation}
It is evident that $\partial^{2}\equiv \eta^{\mu \nu}\partial_{\mu}\partial_{\nu}$ is invariant under a Lorentz transformation.

    Vectors that transform like $x^{\mu}$ in eq. (182) are called ``contravariant''; those that transform like $\partial_{\mu}$ in eq. (186) are called ``covariant''. (These types of vectors can be generalized to deal with transformations more general than the linear Lorentz transformations considered here.) A general tensor $T^{\mu_{1}...\mu_{m}}_{\nu_{1}...\nu_{n}}$ which has $m$ contravariant and $n$ covariant indices transforms as follows
\begin{equation}
T^{'\mu_{1}...\mu_{m}}_{\nu_{1}...\nu_{n}}=\Lambda^{\mu_{1}}_{\;\;\;\alpha_{1}}...\Lambda^{\mu_{m}}_{\;\;\;\alpha_{m}}\Lambda_{\nu_{1}}^{\;\;\;\beta_{1}}...\Lambda_{\nu_{n}}^{\;\;\;\beta_{n}}T^{\alpha_{1}...\alpha_{m}}_{\beta_{1}...\beta_{n}}.
\end{equation}
From eqs. (183) and (184), it follows that 
\begin{equation}
\Lambda_{\mu}^{\;\;\;\nu}=\eta_{\mu \alpha}\eta^{\nu \beta}\Lambda^{\alpha}_{\;\;\;\beta}
\end{equation}
and hence $T^{\mu}_{\mu}$ is a scalar. Furthermore, if $T^{\mu}$ is a contravariant vector, then $T_{\mu}=\eta_{\mu \nu}T^{\nu}$ is a covariant vector.

    It was noted by Thomas that two successive boosts that are not collinear do not in general constitute a boost, but rather are composed of a boost and a rotation. To see this, consider a boost in the $z$ direction characterized by velocity $(0,0,v)$ followed by a boost in the $y-z$ plane characterized by velocity $(0,\delta v_y,\delta v_z)$ with $\delta v_i$ being small. Two successive applications of eqs. (166) and (167) show that the resulting transformation is composed of a boost characterized by velocity $(0,\delta v_y\sqrt{1-v^2},v)$ and a rotation about the $x$ axis through a small angle $\delta \omega \approx \frac{\delta v_y}{v}\left(1-\sqrt{1-v^2}\right)$.  

We note that the Lorentz transformation of eqs. (166-170) do not immediately provide what an observer ``sees'' when observing an object, as it takes a finite 
amount of time for a light signal emanated by an object to reach the observer.  If the signal were emitted at time $\tau$ and received at time $t$ by an observer at the 
origin, then since the velocity of light is constant to all observers 
\begin{eqnarray}
c(t - \tau) = (\mathrm{distance \;travelled\; by\; the \;light\; signal}\nonumber\\
\qquad \mathrm{in\; the\; observer's \;frame\; of\; reference)}.
\end{eqnarray}
If the object is moving with speed $v$ along the $x$ axis, being at the origin at time $t = 0$, then eq. (189) becomes
\begin{equation}
c\left(t - \frac{x_p}{v}\right) = |x_p|\;.
\end{equation}
If $x_p > 0$ so that the object is receding from the observer, then by eq. (190)
\begin{equation}
x_p = \frac{vt}{1+v/c}
\end{equation}
while if $x_p < 0$ so that the object is approaching the observer 
\begin{equation}
x_p = \frac{vt}{1-v/c}.
\end{equation}
The apparent, or observed, velocity of the particle is $v_p = \frac{dx_p}{dt}$.  From eq. (191), we see that for a receding object, $v_p \rightarrow \frac{c}{2}$ as 
$v \rightarrow c$ while from eq. (192) $v_p \rightarrow \infty$ as $v \rightarrow c$ for an approaching object.

The notion of ``constant accelerations'' should also be considered in the context of special relativity. It obviously cannot mean that the speed of 
an object increases indefinitely, as it does in Galilean relativity, as this would mean that the speed would eventually exceed $c$ which contradicts 
special relativity.  What we do mean by ``constant acceleration'' is that the acceleration of an object is the same according to all observers instantaneously at 
rest with respect to that object.  We can now show that this criterion is met by an object following the trajectory
\begin{equation}
x^2 - c^2 t^2 = \alpha^2 .
\end{equation}
As will be discussed below, this trajectory results from application of a constant force.  From eq. (193) it follows that
\begin{equation} \frac{dx}{dt} = \frac{c^2t}{(\alpha^2 + c^2 t^2)^{1/2}} \equiv v
\end{equation}
and
\begin{equation} \frac{d^2x}{dt^2} = \frac{c^2}{(\alpha^2 + c^2 t^2)^{1/2}} -  \frac{c^4t^2}{(\alpha^2 + c^2 t^2)^{3/2}} \equiv a.
\end{equation}
Together, eqs. (193-195) imply that $a = (c^2/\alpha) ( 1 - v^2/c^2)^{3/2}$. From eq. (194), we see that the object is instantaneously at rest at time $t = 0$; by eq. (195) the acceleration of the object is $c^2/\alpha$ at this time. But by eq. (178), 
the trajectory of this particle according to another observer moving with velocity $v$ along the $x$ axis is
\begin{equation}
x^{\prime 2} - c^2 t^{\prime 2} = \alpha^2 .
\end{equation}
This observer will hence see the object at rest at time $t^\prime = 0$ and at this time the acceleration will also be $c^2/\alpha$. Hence the 	``hyperbolic motion'' described 
by eq. (193) is consistent with the criterion for constant acceleration in special relativity.  We note that from eq. (194), $\frac{dx}{\partial t} \rightarrow \pm c$ and by 
eq. (195) $\frac{d^2x}{\partial t^2} \rightarrow 0$ as $t \rightarrow \pm \infty$. 

If such a uniformly accelerating particle is moving along the branch of the hyperbole 
\begin{equation} x^2_p - c^2 t_p^2 = \alpha^2 \end{equation}
which has $x_p > 0$, then as $\tau = \pm \frac{1}{c} \sqrt{x_p^2 - \alpha^2}$, eq. (189) becomes
\begin{equation} c\left[ t - \left( \pm \frac{1}{c} \sqrt{x_p^2 - \alpha^2}\right)\right] = x_p\end{equation}
where the positive sign in eq. (198) is for a receding particle and the negative sign is for an approaching particle.  Solving eq. (198) we find that 
\begin{equation} x_p = \frac{ct}{2} + \frac{\alpha^2}{2ct} .\end{equation}
This is valid only for $x_p > 0$ and $t > 0$; if $t < 0$ then no signal coming from the object can be detected by this observer who is located at the 
origin.  The apparent velocity and acceleration of the particle are $\frac{c}{2} - \frac{\alpha^2}{2ct^2}$ and $\frac{\alpha^2}{ct^3}$ respectively.  This is consistent 
with eqs. (191) and (192), as the limit $t \rightarrow \infty$ corresponds to the object receding from the origin and the signal being emitted at a time 
$\tau \rightarrow \infty$, while $t \rightarrow 0^+$ corresponds to the object approaching the origin and the signal being emitted at a time $\tau \rightarrow -\infty$. 
At times $\tau \rightarrow \pm \infty$, the magnitude of the velocity of the object approaches $c$.

\subsection*{\large 1.7 The Point Particle}

    We now consider the canonical formalism for a free particle. If a particle moves along a space-time trajectory specified by a vector $x^{\mu}(\lambda )$ with specified end points $x^{\mu}(\lambda_i)$ and $x^{\mu}(\lambda_f)$, then the action for determining this trajectory can only be constructed out of the geometrical quantities that characterize this path. The simplest such quantity is just the path length of the trajectory. Since this length is given by $ds^2$ where 
\begin{equation}
ds^2=\eta_{\mu \nu}dx^{\mu}dx^{\nu}
\end{equation}
our initial action is taken to be 
\begin{equation}
S=-m\int^{\lambda_f}_{\lambda_i}\sqrt{ds^2}
\;\;\;=-m\int^{\lambda_f}_{\lambda_i}\sqrt{\dot{x}^{2}}d\lambda
\end{equation}    
where $m$ is a scale parameter (called the ``mass'').  Often $\sqrt{ds^2} = dt\sqrt{1 - \overline{v}^2}$ is called the ``proper time'' as it is an invariant that equals the time 
elapsed between two events that occur in the rest frame of the observer (ie, where $\vec{v} = 0$).

    It is apparent from eq. (201) that the choice of $\lambda$ is arbitrary; one could replace $\lambda$ by $\lambda^{'}=f(\lambda )$ and leave the form of $S$ in eq. (201) unaltered. This form of ``gauge invariance'' is associated with a first class constraint arising in the canonical treatment of $S$. To see this, we note that      
\begin{equation}
p_{\mu}=-m\frac{\partial \sqrt{\dot{x}^2}}{\partial \dot{x}^\mu }=\frac{-m\dot{x}_\mu}{\sqrt{\dot{x}^2}}.
\end{equation}
From eq. (202) we see that not all components of $p_{\mu}$ are independent as 
\begin{equation}
p^2=m^2;
\end{equation}
this is a first class primary constraint. There are no further constraints associated with this action. The canonical Hamiltonian of eq. (13) can easily be shown to vanish and thus by eq. (81) the Hamiltonian is entirely given by the constraint of eq. (203)
\begin{equation}
H=u(p^2-m^2).
\end{equation}
The ``time'' derivative of $x^{\mu}(\lambda )$ (that is, the derivative with respect to $\lambda$) can be found from eq. (94) and thus we get
\begin{equation}
\dot{x}^{\mu}(\lambda )=2up^{\mu}
\end{equation}
as well as 
\begin{equation}
\dot{p}_{\mu}=0.
\end{equation}
From eq. (205) it is apparent that since $u$ is arbitrary, $\dot{x}^{\mu}$ is arbitrary; this is a reflexion of the fact that $\lambda$ itself is arbitrary. 

    The canonical formalism with the gauge choice $x^0=\lambda$ can be pursued, but it is much easier to simply insert this choice of $\lambda$ into the action of eq. (201). This leads to 
\begin{equation}
S=-m\int^{t_f}_{t_i}\sqrt{1-\textbf{v}^2}dt\equiv -m\int \gamma^{-1} dt
\end{equation}    
where we have set $x^{\mu}=(t,\textbf{r})$ and $\textbf{v}=\dot{\textbf{r}}$. With this, we find that 
\begin{equation}
\textbf{p}=\frac{\partial (-m\gamma ^{-1})}{\partial \textbf{v}}=m\textbf{v}\gamma
\end{equation}
and the Hamiltonian becomes
\begin{equation}
H=\textbf{p}\textbf{v}-L=m\gamma .
\end{equation}
(Inserting factors of $c$ in eq. (209) when $\textbf{v}=0$ leads to the famous equation $E=mc^2$.)
From eqs. (207, 208) we see that 
\begin{equation}
H=\sqrt{\textbf{p}^2+m^2}.
\end{equation}

    If we supplement the action of eq. (207) with a potential associated with a force $F$ in the spatial direction $x$, we have the Lagrangian $L=-m\sqrt{1-\dot{x}^2}+Fx$; eq. (3) then yields the equation of motion 
\begin{equation}
\frac{d\frac{m\dot{x}}{\sqrt{1-\dot{x}^2}}}{dt}=F.
\end{equation}
We see immediately that $m\ddot{x}=F$ when $\dot{x}=0$, that is, in the rest frame of the particle, we have Newton's equation $\textbf{F}=m\textbf{a}$. If $x=\dot{x}=0$ when $t=0$, then eq. (211) can be integrated to yield the trajectory given in eq. (193),
\begin{equation}
t^2-x^2=-\left(\frac{m}{F}\right)^2.
\end{equation}
Thus we see that a constant force results in ``hyperbolic motion'' in Minkowski space.

    The action of eq. (201) disappears in the limit $m=0$. Furthermore, the expressions for momentum and energy given by eqs. (208) and (209) respectively would vanish if $\textbf{v}^2$ were not to approach $1$. In order to handle this massless limit, we introduce an auxiliary field $e$ and consider the Lagrangian
\begin{equation}
L=-\frac{1}{2}\left(\frac{\dot{x}^2}{e}+m^2e\right).
\end{equation}
The equation of motion for $e$ is just 
\begin{equation}
e=\frac{\sqrt{\dot{x}^2}}{m}
\end{equation}
which, when substituted back into eq. (213) just leads to the Lagrangian appearing in eq. (201). The advantage of introducing $e$ is that it is possible to let $m$ vanish in eq. (213) and retain realistic dynamics; it is apparent that in this massless limit the equation of motion for $e$ implies that $\dot{x}^2=0$ so that massless particles move at the speed of light. If one were to make a canonical analysis of eq. (213), one encounters the primary constraint $p_e=0$ for the momentum conjugate to $e$ while for the momentum conjugate to $x^\mu$ is $p_{\mu} =-\frac{\dot{x}_{\mu}}{e}$. The Hamiltonian becomes  
\begin{equation}
H=\frac{e}{2}\left(-p^2+m^2\right)
\end{equation}
so that requiring that $\dot{p}_e=0$ generates the secondary first class constraint of eq. (203). The primary constraint $p_e=0$ is also first class. With $\pi 1=p_e$ and $\pi 2=\frac{1}{2}(p^2-m^2)$, eqs. (118) and (119) lead to a gauge transformation generator 
\begin{equation}
G=\frac{1}{2}\epsilon \left(-p^2+m^2\right)+\dot{\epsilon}p_e.
\end{equation}
This results in the infinitesmal changes
\begin{equation}
\delta e=\left\{G,e\right\}=-\dot{\epsilon}
\end{equation}
and
\begin{equation}
\delta x^\mu =\left\{G,x^\mu \right\}=\epsilon p^\mu =-\frac{\epsilon \dot{x}^\mu}{e}.
\end{equation}
The transformations of eqs. (217,218) leave the action associated with the Lagrangian $L$ of eq. (213) unaltered as $L$ changes only by a total derivative,
\begin{equation} 
\delta L=\frac{1}{2}\frac{d}{d\lambda}\left[\left(\frac{\dot{x}^2}{e}+m^2\right)\epsilon \right]. 
\end{equation}
 
    Arc length, used in eq. (201), is not the only quantity that can be employed to form an action for a particle moving along a trajectory $x^\mu (\lambda )$; one can also use the extrinsic curvature of the trajectory. The action of eq. (201) is then modified to become 
\begin{equation}
S=-\int ^{\lambda_f}_{\lambda_i}ds\left(m-\mu \left(\frac{d^2x^\mu}{ds^2}\right)^2\right)
\end{equation}  
where again $ds$ denotes the arc length $\sqrt{dx^{\mu}dx_{\mu}}$. This action does not possess the reparameterization invariance 
of eq. (201).  If we now express $x^\mu$ as a function of $t=x^0$ as in eq. (207), then
\begin{equation}
\frac{d^2x^\mu}{ds^2}=\left(\frac{\textbf{v}.\textbf{a}}{\left(1-\textbf{v}^2\right)^2},\frac{\textbf{a}}{1-\textbf{v}^2}+\frac{\textbf{v}\textbf{a}.\textbf{v}}{\left(1-\textbf{v}\right)^2}\right)
\end{equation}
where $\textbf{a}\equiv \frac{d\textbf{v}}{dt}$ and $\textbf{v}\equiv \frac{d\textbf {x}}{dt}$. The action itself is then 
\begin{equation}
S=-\int ^{t_f}_{t_i}\left(m\gamma ^{-1}+\mu \left(\textbf{a}^2\gamma^3+\left(\textbf{a}.\textbf{v}\right)^2\gamma ^5\right)\right)dt.
\end{equation}
Since this action depends on $\textbf{a}=\ddot{\textbf{x}}$, the equation of motion is given by eq. (11) and we have two canonical momenta
\begin{equation}
\pi =\frac{\partial L}{\partial \textbf{a}}= -2\mu \left(\gamma^3 \textbf{a}+\gamma^5 \textbf{v}\textbf{a}.\textbf{v}\right)
\end{equation}
and
\begin{equation}
\textbf{p}=\frac{\partial L}{\partial \textbf{v}}-\frac{d}{dt}\frac{\partial L}{\partial \textbf{a}}.
\end{equation}
From eq. (223) it follows that 
\begin{equation}
\textbf{a}=-\left(2\mu \gamma^3\right)^{-1}\left(\pi -\textbf{v}\pi .\textbf{v}\right).
\end{equation} 
We also have 
\begin{equation}
\textbf{v}=\dot{\textbf{x}}.
\end{equation}
The Hamiltonian is then given by 
\begin{equation}
H\left(\textbf{x},\textbf{v};\textbf{p},\pi \right)=\textbf{p}.\dot{\textbf{x}}+\pi .\dot{\textbf{v}}-L\left(\textbf{x},\dot{\textbf{x}},\ddot{\textbf{x}}\right),
\end{equation}
which becomes
\begin{equation}
H=\textbf{p}.\textbf{v}+m\gamma^{-1}-\left(2\mu\gamma^3\right)^{-1}\left(\pi^2-\left(\textbf{v}.\pi\right)^2\right).
\end{equation}
The equations of motion for any quantity $A$ are 
\begin{equation}
\frac{dA}{dt}=\left\{A,H\right\}
\end{equation}
where now the Poisson bracket is given by
\begin{equation}
\left\{A,B\right\}=\frac{\partial A}{\partial \textbf{x}}.\frac{\partial B}{\partial \textbf{p}}-\frac{\partial A}{\partial \textbf{p}}.\frac{\partial B}{\partial \textbf{x}}+\frac{\partial A}{\partial \textbf{v}}.\frac{\partial B}{\partial \pi}-\frac{\partial A}{\partial \pi}.\frac{\partial B}{\partial \textbf{v}}.
\end{equation}
With the Hamiltonian of eq. (228), it follows that $\textbf{L}+\textbf{S}$ is conserved, where $\textbf{L}$, the ``orbital'' angular momentum, is given by $L_{i}=\epsilon_{ijk}x_{j}p_{k}$ and $\textbf{S}$, the ``spin'' angular momentum, is given by $S_{i}=\epsilon_{ijk}v_{j}\pi_{k}$ . 

    If circular motion is postulated, so that 
\begin{equation}
\textbf{x}=R\left(cos\left(\omega t\right),sin\left(\omega t\right),0\right)
\end{equation}    
in a ``rest'' frame, then by eqs. (171-173) in a frame boosted with velocity $u$ in a direction perpendicular to the plane of rotation,
\begin{equation}
\dot{\textbf{x}}=\left(-v\sqrt{1-u^2}  sin(\omega t),v\sqrt{1-u^2}  cos(\omega t),u\right)
\end{equation}
where $v\equiv \omega R$.
Together, eqs. (228-230) lead to 
\begin{equation}
\dot{\textbf{x}}=\textbf{v}
\end{equation}
\begin{equation}
\dot{\textbf{p}}=0
\end{equation}
\begin{equation}
\dot{\textbf{v}}=-\left(\mu \gamma^3\right)^{-1}\left(\pi-\textbf{v}\pi .\textbf{v}\right)
\end{equation}
and
\begin{equation}
\dot{\pi}=-\frac{3}{2\mu \gamma}\left(\pi^2-\left(\pi .\textbf{v}\right)^2\right)\textbf{v}-\frac{1}{\mu \gamma^3}\pi .\textbf{v}\pi -\textbf{p}+m\gamma \textbf{v}.
\end{equation}
With eq. (232), eqs. (233-236) show that 
\begin{equation}
\pi=\frac{\omega v\mu}{\alpha^{\frac{3}{2}}\beta}\left(cos(\omega t),sin(\omega t),0\right)
\end{equation}
where $\alpha\equiv 1-v^2$ and $\beta\equiv 1-u^2$. Furthermore, it follows that
\begin{equation}
\textbf{p}=\left(0,0,\frac{mu}{\sqrt{1-u^2}}\frac{\sqrt{1-v^2}}{1+\frac{v^2}{2}}\right)
\end{equation}
provided that 
\begin{equation}
\omega^2=\frac{2m}{\mu} \frac{\alpha^2\beta}{3-\alpha}.
\end{equation}
The numerical value of the Hamiltonian in eq. (228) is 
\begin{equation}
E=\frac{m}{\sqrt{1-u^2}}\frac{\sqrt{1-v^2}}{1+\frac{v^2}{2}},
\end{equation}
and the total angular momentum is 
\begin{equation}
\textbf{L}+\textbf{S}=\left(0,0,\frac{v^2\sqrt{2m\mu }}{\sqrt{2+v^2}\sqrt{1-v^2}}\right)
\end{equation}
with $\textbf{L}$ and $\textbf{S}$ not being separately conserved.

   The above discussion shows that by incorporating the extrinsic curvature of the trajectory of a particle into its Lagrangian, some qualitative features of the intrinsic spin of a particle can be recovered. However, a more satisfactory classical realization of spin is obtained by making use of Grassmann coordinates. If $\chi (\lambda )$ and $\psi^{\mu} (\lambda )$ are Grassmann variables, then the massless limit of eq. (213) can be generalized to 
\begin{equation}
L=-\frac{1}{2}\left(\frac{\dot{x}^2}{e}+i\psi .\dot{\psi}-i\chi \dot{x}.\psi \right).
\end{equation} 
If $\alpha (\lambda )$ is an arbitrary Grassmann function, then the Lagrangian of eq. (242) undergoes a change 
\begin{equation}
\delta L=\frac{d}{d\lambda }\left(\frac{-i\alpha \dot{x}.\psi}{2e}\right)
\end{equation}
when
\begin{equation}
\delta x^{\mu}=i\alpha \psi^{\mu},
\end{equation}
\begin{equation}
\delta e=-i\alpha \chi,
\end{equation}
\begin{equation}
\delta \psi^{\mu}=-\alpha \left(\frac{\dot{x}^{\mu}}{e}-\frac{i\chi \psi^{\mu}}{2e}\right)
\end{equation}
and
\begin{equation}
\delta \chi=2\dot{\alpha}
\end{equation}
and consequently the action itself is invariant.

    The momenta associated with $e$ and $\chi$ vanish; these are first class constraints which can be associated with the gauge conditions $\chi =0$ and $e=-1$. The equations of motion then collapse down to the dynamical equations 
\begin{equation}
\ddot{x}=\dot{\psi} =0
\end{equation}
as well as the constraint equations
\begin{equation}
\dot{x}^2=\dot{x}.\psi =0.
\end{equation}    
These follow from the Lagrangian
\begin{equation}
L=\frac{1}{2}\left(\dot{x}^2-i\psi .\dot{\psi}\right)
\end{equation}
which shows that the momentum associated with $\psi^{\mu}$ is proportional to $\psi^{\mu}$ itself. This is a second class constraint whose Dirac bracket, determined from eqs. (91) and (151), imply that
\begin{equation}
\left\{\psi^{\mu},\psi^{\nu}\right\}^{*}= i\eta^{\mu \nu}
\end{equation}
An extension of eq. (242) to the case in which $m^2\neq 0$ requires an additional Grassmann field $\psi^5$. The Lagrangian becomes 
\begin{equation}
L=-\frac{1}{2}\left[\frac{\dot{x}^2}{e}+m^2e+i\left(\psi .\dot{\psi} -\psi^5 \dot{\psi}^5\right) -i\chi \left(\frac{\psi .\dot{x}}{e}+m\psi^5\right)\right].
\end{equation}
Now the dynamical equations of motion are
\begin{equation}
\ddot{x}^{\mu}=\dot{\psi}^{\mu}=\dot{\psi}^{5}=0.
\end{equation}

The constraint formalism can be applied.  The canonical momenta are 
\begin{subequations}
\begin{align}
p_\mu &= \frac{\partial L}{\partial \dot{x}^\mu} = \frac{-1}{e} \left( \dot{x}_\mu - \frac{i}{2} \chi\psi_\mu \right)\\
p_e &= \frac{\partial L}{\partial \dot{e}} = 0 \\
\pi_\mu &= \frac{\partial L}{\partial \dot{\psi}^\mu} = \frac{i}{2} \psi_\mu \\
\pi_5 &= \frac{\partial L}{\partial \dot{\psi}_5} = -\frac{i}{2} \psi_5 \\
\pi_\chi &= \frac{\partial L}{\partial \dot{\chi}} = 0,
\end{align}
\end{subequations}
so that the canonical Hamiltonian is
\begin{equation}
H_C = -\frac{e}{2} (p^2 - m^2) + \frac{i}{2} \chi (p \cdot\psi - m \psi_5).
\end{equation}
With the primary second class constraints of eqs. (254c,d) we obtain the Dirac Brackets (see eq. (91))
\begin{align}
\left\lbrace A,B \right\rbrace^* &= \left\lbrace A,B \right\rbrace + i \bigg[ \left\lbrace A, \pi_\mu - \frac{i}{2} \psi_\mu \right\rbrace \left\lbrace \pi^\mu - \frac{i}{2} \psi^\mu , B \right\rbrace \\
\hspace{2cm} &- \left\lbrace A, \pi_5 + \frac{i}{2} \psi_5\right\rbrace \left\lbrace \pi_5 + \frac{i}{2} \psi_5, B\right\rbrace \bigg]\nonumber
\end{align}
from which it follows that 
\begin{subequations}
\begin{align}
\left\lbrace \psi_\mu , \psi_\nu \right\rbrace^* &= i\; \eta_{\mu\nu} \\
\left\lbrace \psi_5 , \psi_5 \right\rbrace^* &= -i .
\end{align}
\end{subequations}
The primary constraints of eqs. (254b,e) yield secondary constraints as 
\begin{subequations}
\begin{align}
\left\lbrace p_e, H_C \right\rbrace &= \frac{1}{2} (p^2 - m^2) \\
\left\lbrace \pi_\chi , H_C \right\rbrace &= -\frac{i}{2}  (p \cdot \psi - m\psi_5)
\end{align}
\end{subequations}
which are first class as 
\begin{equation}
\left\lbrace p \cdot \psi - m \psi_5, \; p \cdot \psi - m \psi_5 \right\rbrace^* = i (p^2 - m^2); 
\end{equation}
there are no tertiary (third generation) constraints.

The gauge generator of eq. (130) is now 
\begin{equation}
G = \mu^a \phi_a = B_1 p_e + B(p^2 - m^2) + i F_1 \pi_\chi + i F((p \cdot \psi - m\psi_5)
\end{equation}
where $(B, B_1)$ and $(F, F_1)$ are Bosonic and Fermionic gauge functions respectively.  Taking the total Hamiltonian to be
\begin{align}
H_T& = H_c + U^{\theta_{1}} \phi_{a_{1}} \nonumber \\
&= -\frac{e}{2} (p^2 - m^2) + \frac{i}{2} \chi (p \cdot \psi - m\psi_5) + \lambda_e p_e + i\lambda_\chi \pi_\chi
\end{align}
then eq. (131) becomes
\begin{align}
\dot{B}_1 p_e + \dot{B} (p^2 - m^2) &+ i \dot{F}_1 \pi_\chi + i \dot{F}(p \cdot \psi - m \psi_5) \nonumber \\
&+ \left\lbrace G, H_T \right\rbrace^* - \delta \lambda_a p_e - i\delta\lambda_\chi \pi_\chi = 0.
\end{align}
(We use the Dirac Bracket in eq. (262) which permits us to set the second class constraints equal to zero at the outset.) From eq. (262) we find that 
\begin{subequations}
\begin{align}
B_1 &= 2\dot{B} + i F\chi\\
F_1 &= -2i\dot{F} .
\end{align}
\end{subequations}
Taking $G$ to be given by eqs. (260, 263), then the change $\delta A = \left\lbrace A, G \right\rbrace^*$ leads to 
\begin{subequations}
\begin{align}
\delta\chi^\mu &= 2Bp^\mu + iF\psi^\mu = -\frac{2B}{e} \left( \dot{x}^\mu - \frac{i}{2} \chi\psi^\mu\right) + iF\psi^\mu\\
\delta e&= 2\dot{B} + iF\chi \\
\delta\chi &= 2\dot{F}\\
\delta\psi_\mu &= Fp^\mu = -\frac{F}{2} \left( \dot{x}^\mu - \frac{i}{2} \chi\psi^\mu \right)\\
\delta\psi_5 &= mF.
\end{align}
\end{subequations}
These transformations have a group property determined by 
\begin{equation}
\left\lbrace G(B_a, F_a), \, G(B_b, F_b)\right\rbrace^* =2 \frac{d}{d\tau} (i F_aF_b) p_e + (iF_aF_b)(p^2 -m^2).
\end{equation}
We see that commuting two gauge transformations generated by $G$ leads to a pure Bosonic gauge transformation with gauge generator $B = iF_aF_b$.

\end{document}